\documentclass[11pt,onecolumn,conference]{IEEEtran}

\usepackage{graphicx}  
\begin{document}

\pagestyle{plain}


\title{
Parallel Computing for 4-atomic Molecular Dynamics Calculations}
\author{\authorblockN{Renat A. Sultanov\authorrefmark{1}, Mark Nordby and Dennis Guster}
\authorblockA{Business Computer Research Laboratory, St. Cloud State University,\\
BB-252, 720 Fourth Avenue South, St Cloud, MN 56301-4498}
$^*$Email: rasultanov@stcloudstate.edu; r.sultanov2@yahoo.com\\
}


%


\maketitle
\begin{abstract}
We report the results of intensive numerical calculations for four atomic H$_2$+H$_2$ energy transfer collision.
A parallel computing technique based on LAM/MPI functions is used. In this algorithm, the data is
distributed to the processors according to the value of the momentum quantum number $J$ and its projection $M$.
Most of the work is local to each processor. The topology of the data communication is a simple star.
Timings are given and the scaling of the algorithm is discussed. 
Two different recently published potential energy surfaces for the H$_2-$H$_2$ system are applied.
New results obtained for the state resolved excitation-deexcitation cross sections and rates valuable 
for astrophysical applications are presented. Finally, more sophisticated extensions of the parallel code are
discussed.
%
%
\end{abstract}
\vspace{2mm}
\hspace{8mm}{\bf Keywords}: Parallel algorithm, LAM/MPI application, Star-type cluster, quantum dynamics.
\vspace{5mm}
%
\IEEEpeerreviewmaketitle

\section{Introduction}

%
%
%

In modern competitive research in science and technology high performance computing 
plays a paramount role. Its importance is derived from the fact, that correctly chosen and designed numerical methods
and algorithms properly adapted to parallel and multithreaded techniques
can essentially reduce computation time and active memory usage \cite{medved05}.
The importance of this fact is especially magnified in calculating quantum molecular dynamics and atomic collisions 
due to their massive complexity.              


Generally speaking, modern computation research in scientific applications has taken two twists. First,
to provide efficient and stable numerical calculations, and second to provide for the
proper use of various high performance techniques like LAM/MPI, OpenMP and/or 
others \cite{openmp}. Now it is equally important not only to get the correct numerical results, but also to design and 
implement
efficient high performance algorithms and get faster results with less memory.
We would like to note here, that a program/software, which is designed for specific problems
in computational physics, chemistry or biology should be able to perform calculations in either serial or 
parallel.

The problem we selected for our
parallel computation in this work is taken from molecular/chemical physics.
Specifically we carry out {\it detailed} quantum-mechanical calculations of state-resolved cross sections and rates
in hydrogen molecular collisions H$_2$+H$_2$. 
Interaction and collisions between hydrogen molecules, and hydrogen molecular isotopes, for example H$_2$+HD,
is of great theoretical and experimental interest for many years [3-14].
Specifically we will explore the 
quantum-mechanical 4-atomic system shown in Fig.\ 1 using six independent variables resulting in the
full description of the system. The main goal of this investigation is to carry out a comparative analysis of two 
recently published potential energy surfaces (PESs) for H$_2-$H$_2$.

Our motivation for selecting this problem is, that the hydrogen molecule plays an important
role in many areas of astrophysics [15-16] 
This is the simplest and most abundant 
molecule in the universe especially in giant molecular clouds. 
Because of low number of electrons in H$_2-$H$_2$ this is one of few four-center systems for which 
potential energy surface (PES) can be
developed with very high precision. Therefore H$_2$+H$_2$ may be also a benchmark collision for testing other
dynamical methods. Additionally, the H$_2$+H$_2$ elastic and inelastic collisions are
of interest in combustion, spacecraft modeling and at the present hydrogen gas is becoming a 
very important potential energy supplier, see for example \cite{zuttel04}.

We test two PESs: the first one is a global 6-dimensional potential from work
\cite{booth02}, the second one is very accurate interaction potential calculated from the first principles
\cite{diep00}. Because we are going to carry out detailed quantum-mechanical
calculations using two PESs the computation work is at least doubled and therefore even more time consuming.
We needed to carry out convergence tests with respect to different chemical and numerical parameters
for both PESs and, finally, we have to make production calculations for many points of kinetic 
energy collisions. Clearly, an application of parallel computing techniques shall be very useful in this situation.

%
%

In this work we carry out parallel computation with up to 14 processors.
The scattering cross sections and their
corresponding rate coefficients are calculated using a non reactive quantum-mechanical close-coupling
approach. In the next section we will shortly outline the quantum-mechanical method and the parallelization approach.
Our calculations for H$_2$+H$_2$, scaling and timing results are presented in Sec. III.
Conclusions are given in Sec. IV. Atomic units (e=m$_e$=$\hbar$=1) are used throughout the work.

\begin{figure}
\begin{picture}(250,250)(-30,0)
\put(160,20){\circle*{23}}
\put(153,0){\sf{(a)}}

\put(135,20){\bf H}
\put(265,-6){\circle*{23}}
\put(258,-26){\sf{(b)}}

\put(280,-6){\bf H}
\put(160,20){\vector(4,-1){98}}
\put(245,3){$\vec r_1$}
\put(212.5,7){\vector(0,1){165}}
\put(200,150){$\vec R$}
\put(212.5,-20){\line(0,1){240}}
\put(200,220){$Z$}
\put(200,-3){$O$}
\put(215,8){$\Theta_1$}
\put(215,185){$\Theta_2$}

\qbezier[60](200,100)(260,108)(205,110)
\put(205,110){\vector(-1,0){4}}
\put(215,115){$\Phi_2$}

\put(165,150){\circle*{23}}
\put(158,131){\sf{(c)}}
\put(135,150){\bf H}

\put(260,195){\circle*{23}}
\put(253,176){\sf{(d)}}
\put(280,195){\bf H}

\put(165,150){\vector(2,1){88}}
\put(240,200){$\vec r_2$}
\end{picture}

\vspace{10mm}

\caption{4-body coordinates for the H$_2-$H$_2$ system used in this work.}
\label{fig:fig1}
\end{figure}
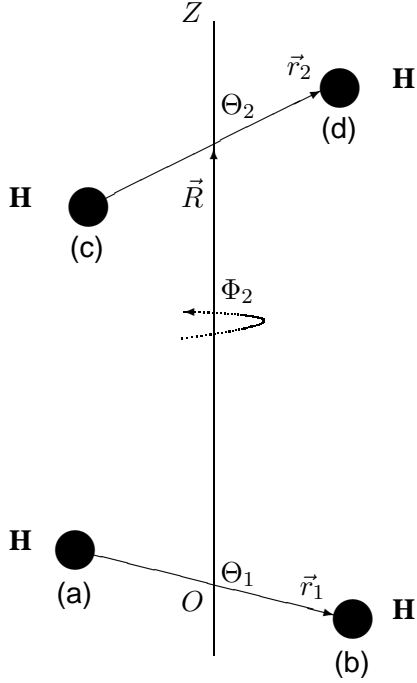


\section{Method}

\subsection{Quantum-mechanical approach}

In this section we briefly represent a quantum-mechanical approach and the parallel algorithm used in this work.
The 4-atomic H$_2-$H$_2$ system is shown in Fig. 1. It can be described by six independent variables:
$r_{1}$ and $r_{2}$ are interatomic distances in each hydrogen
molecule,  $\Theta_{1}$ and $\Theta_{2}$ are polar angles,  $\Phi$ is torsional angle and $R$ is intermolecule distance.
%
%
The hydrogen molecules are treated as 
linear rigid rotors, that is distances $r_1=r_2=0.74 A$ are fixed in this model.
We provide a numerical solution for the Schr\"odinger equation for an $ab+cd$ collision in the center of the
mass frame, where $ab$ and $cd$ are linear rigid rotors.

%
%
%
%
%
%
%
%
%
To solve the equation the total 4-atomic H$_2+$H$_2$ wave function is expanded into channel angular momentum functions 
$\phi^{JM}_{\alpha'}(\hat r_1,\hat r_2,\vec R)$ \cite{green75}.
%
%
%
%
%
%
This procedure followed by separation of angular momentum    
provides a set of coupled second order differential equations for the unknown radial functions $U^{JM}_{\alpha}(R)$
\begin{eqnarray}
\left(\frac{d^2}{dR^2}-\frac{L(L+1)}{R^2}+k_{\alpha}^2\right)U_{\alpha}^{JM}(R)=2M_{12}
\sum_{\alpha'} \int <\phi^{JM}_{\alpha}(\hat r_1,\hat r_2,\vec R)\nonumber \\
|V(\vec r_1,\vec r_2,\vec R)| 
\phi^{JM}_{\alpha'}(\hat r_1,\hat r_2,\vec R)>U_{\alpha'}^{JM}(R) d\hat r_1 d\hat r_2 d\hat R,
\label{eq:cpld}
\end{eqnarray}
where $\alpha \equiv (j_1j_2j_{12}L)$, $j_1+j_2=j_{12}$, $j_{12}+L=J$ and $j_1, j_2, L$ are quantum angular momentum
corresponding to vectors $\vec r_1$, $\vec r_2$ and $\vec R$ respectively, 
$M_{12} = (m_a+m_b)(m_c+m_d)/(m_a+m_b+m_c+m_d)$, 
$V(\vec r_1,\vec r_2,\vec R)$ is the potential energy surface for the 4-atomic system $abcd$, and $k_{\alpha}$ is
channel wavenumber.

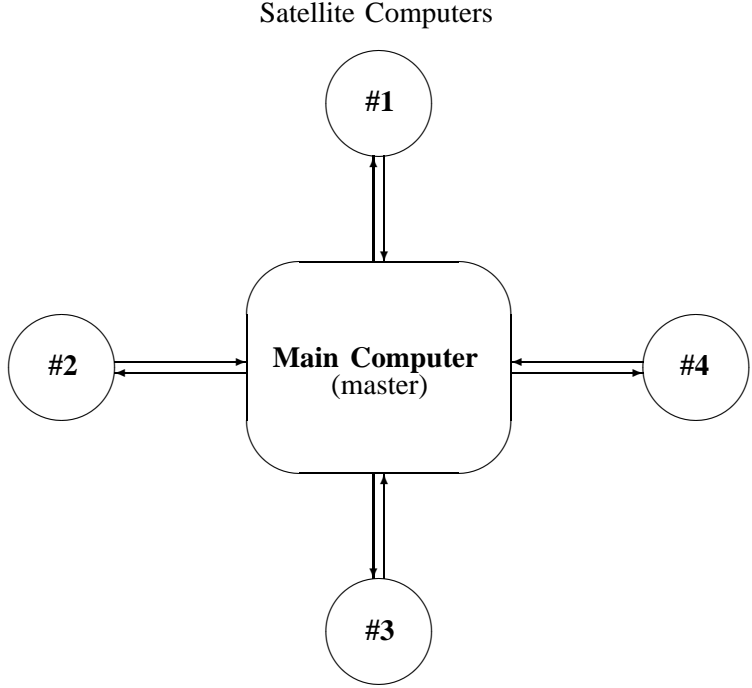
\begin{figure}
\begin{picture}(250,250) 

\put(250,125){\oval(100,80)}

\put(250,225){\circle{40}}
\put(250,25){\circle{40}}
\put(130,125){\circle{40}}
\put(370,125){\circle{40}}

\put(248,165){\vector(0,1){40}}
\put(252,205){\vector(0,-1){40}}

\put(248,85){\vector(0,-1){40}}
\put(252,45){\vector(0,1){40}}

\put(200,123){\vector(-1,0){50}}
\put(150,127){\vector(1,0){50}}

\put(300,123){\vector(1,0){50}}
\put(350,127){\vector(-1,0){50}}

\put(245,222){\bf \#1}
\put(205,256){Satellite Computers}

\put(245,22){\bf \#3}
\put(125,122){\bf \#2}
\put(364,122){\bf \#4}

\put(210,125){\bf Main Computer}
\put(232,115){(master)}

\end{picture}

\vspace{3mm}

\caption{Schematic diagram of the topology of interprocessor communication: 
an example of a star-type cluster with one main and four satellite computers.}
\label{fig:fig2}
\end{figure}
\vspace{3mm}

We apply the hybrid modified log-derivative-Airy propagator in the general purpose scattering code MOLSCAT 
\cite{hutson94} to solve the coupled radial equations (\ref{eq:cpld}). We have tested other
propagator schemes included in the code. Our calculations showed that other propagators are also
quite stable for both the H$_2-$H$_2$ potentials considered in this work.

Since all experimentally observable
quantum information about the collision is contained in the asymptotic behaviour of functions 
$U^{JM}_{\alpha}(R\rightarrow\infty)$, the log-derivative matrix is propagated to large $R$-intermolecular 
distances. The numerical results are matched to the known asymptotic solution to 
derive the physical scattering $S$-matrix \cite{green75}.
%
%
%
The method was used for each partial wave until a converged cross section was obtained. 
It was verified that results are converged with respect to the number of partial waves as well as
the matching radius $R_{max}$ for all channels included in the calculations.
Cross sections for rotational excitation and relaxation phenomena can be obtained directly from the $S$-matrix.
In particular the cross sections for excitation from $j_1j_2\rightarrow j'_1j'_2$ summed over final $m'_1m'_2$
and averaged over initial $m_1m_2$ are given by
\begin{eqnarray}
\sigma(j'_1,j'_2;j_1j_2,\epsilon)=  
{\pi}/{(2j_1+1)(2j_2+1)k_{\alpha\alpha'}}
\sum_{Jj_{12}j'_{12}LL'}(2J+1)|\delta_{\alpha\alpha'}-   
S^J(j'_1,j'_2,j'_{12}L';j_1,j_2,j_{12},L; E)|^2.
\label{eq:cross}
\end{eqnarray}
The kinetic energy is 
$
\epsilon=E-B_1j_1(j_1+1)-B_2j_2(j_2+1),
$
where $B_{1(2)}$ are rotation constants of rigid rotors $ab$ and $cd$ respectively.

The relationship between a rate coefficient $k_{j_1j_2\rightarrow j'_1j'_2}(T)$ and the corresponding
cross section $\sigma_{j_1j_2\rightarrow j'_1j'_2}(E_{kin})$ can be obtained through the following
weighted average
\begin{equation}
k_{j_1j_2\rightarrow j'_1j'_2}(T) = \frac{8k_BT}{\pi\mu}\frac{1}{(k_BT)^2}\int_{\epsilon_s}^{\infty}
\sigma_{j_1j_2\rightarrow j'_1j'_2}(\epsilon)e^{-\epsilon/k_BT}\epsilon d\epsilon,
\label{eq:rate}
\end{equation}
where 
$T$ is temperature, $k_B$ is Boltzmann constant, $\mu$ is reduced mass of the molecule-molecule system,
and $\epsilon_s$ is the minimum kinetic energy for the levels $j_1$ and $j_2$ to become accessible.


\subsection{Parallelization}

In this work to support parallel computation the following
machines are used: Sun Netra-X1 (UltraAX-i2) with 128 MB RAM (512 MB Swap) and
        500 Mhz UltraSPARC-IIe processor. The master computer is SunFire v440 with 8 GB RAM 
four 1.062 Ghz UltraSPARC-IIIi processors. The system is schematically shown in Fig.\ 2. In this work we apply
LAM/MPI to provide the parallel environment in the cluster.

It is important in the parallel algorithm used in this work, that calculations for specific values of 
$J$ and $M$ are essentially independent. In the PMP MOLSCAT program \cite{pmp2005}, which is used
the parallelization is done over the loop on values $J$ and $M$. The code distributes the required $JM$ pairs across 
the available processors. The computational work distribution is shown schematically in Fig.\ 3.
The same idea has been used in works \cite{renat03,renat03a} for semiquantal atomic collisions.
In these works the parallelization was done along the impact factor $\rho$
of colliding particles, because the solution of the resulting dynamical equations doesn't depend on $\rho$. 
It is well known, that in the semiclassical approach the impact factor $\rho$ is an analog of quantum $J$ number.

\begin{figure}
\begin{picture}(120,120) 

\put(90,40){\vector(1,0){300}}
\put(398,37){$JM$}
\multiput(105,40)(15,0){19}{\line(0,1){5}}
\put(94,35){ \huge{[} }
\put(161,35){ \huge{]} }
\put(168,35){ \huge{[} }
\put(237,35){ \huge{]} }
\put(105,55){$JM$ pairs \#1}
\put(110,10){machine \#1}
\put(179,55){$JM$ pairs \#2}
\put(183,10){machine \#2}
\put(135,23){\vector(0,1){12}}
\put(210,23){\vector(0,1){12}}
\put(270,23){\vector(0,1){12}}
\put(330,23){\vector(0,1){12}}
\put(261,55){\bf{. . .}}
\put(261,10){\bf{. . .}}
\put(285,35){ \huge{[} }
\put(359,35){ \huge{]} }
\put(301,55){$JM$ pairs \#n}
\put(304,10){machine \#n}
\end{picture}

\vspace{1mm}

\caption{{$J/M$-}parallelization method, see text.}
\label{fig:fig3}
\end{figure}
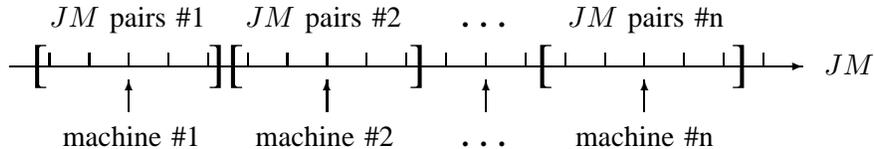

As mentioned above, in the quantum-mechanical approach used in this work, a partial wave expansion is applied.
A set of coupled channel differential equations has to be solved for many values of the total angular momentum
$J$. To calculate the state resolved cross sections and then the rate coefficients (\ref{eq:rate}) the resulting
$S$-matrix elements have to be summed from different $J$s. Calculations for a single $J$ can be broken 
into two or more sectors corresponding to different values of $M$, which is a projection of $J$. 

There are two methods
to distribute the work among satellite computers. In the static method in the beginning of the job
each computer makes a list of the total $J/M$ tasks to be solved. Then each computer selects a subset of the tasks
to carry out. Obviously each computer has to get a different subset and an approach needs to be used which
gives an approximately equal amount of work to each computer.  
There is no interprocessor communication in this method.

In the case of a dynamic approach one computer acts as a dispatcher. It makes a list of all the $J/M$ tasks to be done,
then waits for the computational processes to call in requesting work. 
Starting with the longest tasks, the dispatcher
hands out $J/M$ tasks to computing processes until all of them have been completed.
The next time the computational process
asks for work, the dispatcher sends it a message, and the computational process then does its end-of-run cleanup and exits.

\section{Results}



Our results from the parallel calculations using MPI functions to determine
rotational transitions in collisions between $para/para$- and ortho-/ortho-hydrogen molecules:
\begin{equation}
\mbox{H}_2(j_1) +\mbox{H}_2(j_2) \rightarrow \mbox{H}_2(j'_1) + \mbox{H}_2(j'_2).
\label{eq:h2h2}
\end{equation}
are presented in this section together with scaling results.

As we mentioned in the Introduction we are applying the new PESs from the works \cite{booth02} and  \cite{diep00}.
The DJ PES \cite{diep00} is constructed for
the vibrationally averaged rigid monomer model of the H$_2$$-$H$_2$ system to the complete basis set limit using 
coupled-cluster theory with single, double and triple excitations. A four term spherical harmonics expansion 
model was chosen to fit the surface. It was demonstrated, that
the calculated PES can reproduce the quadrupole moment to within 0.58 \% and 
the experimental well depth to within 1 \%.

\begin{figure}
\begin{center}
\includegraphics[width=25pc,height=18pc]{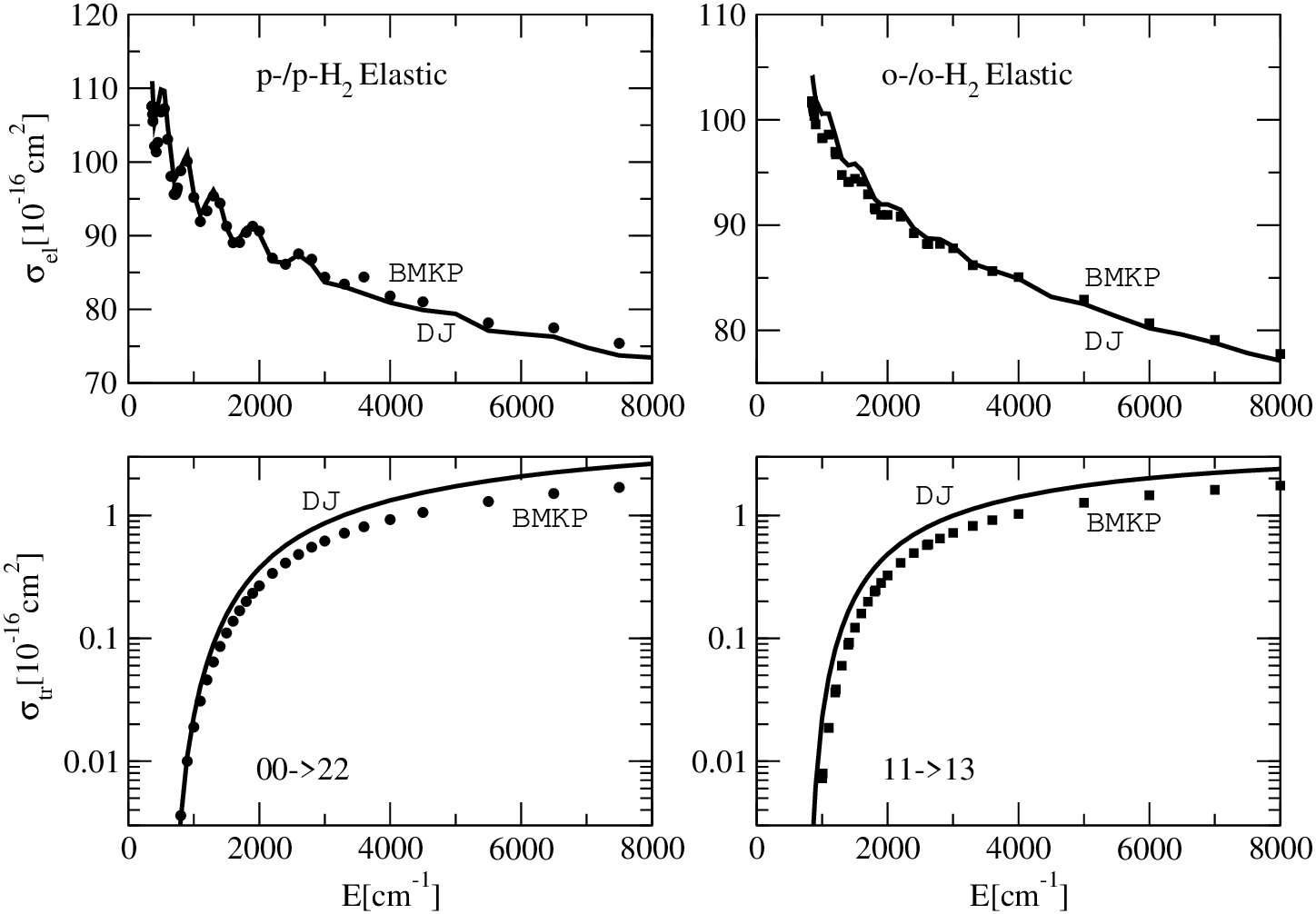}
\end{center}
\caption{Rotational state resolved integral cross sections for elastic scattering in the case of para-/para- and
ortho-/ortho-hydrogen and transitions, when $j_1=j_2=0 \rightarrow j'_1=2,j'_2=2$ and
$j_1=j_2=1 \rightarrow j'_1=1,j'_2=3$. Calculations are done with the BMKP and DJ PESs (the compensating factor 
of 2 is included only in the elastic cross sections).}
\label{fig:fig4}
\end{figure}

The bond length was fixed at 1.449 a.u. or 0.7668 \r{A}. DJ PES is 
defined by the center-of-mass intermolecular distance, $R$, and three angles: $\theta_1$ and $\theta_2$ are the 
plane angles and $\phi_{12}$ is the relative torsional angle. The angular increment for each of the three angles 
defining the
relative orientation of the dimers was chosen to be $30^{\circ}$. 


\begin{figure}
\begin{center}
\includegraphics[scale=1.0,width=23pc,height=17pc]{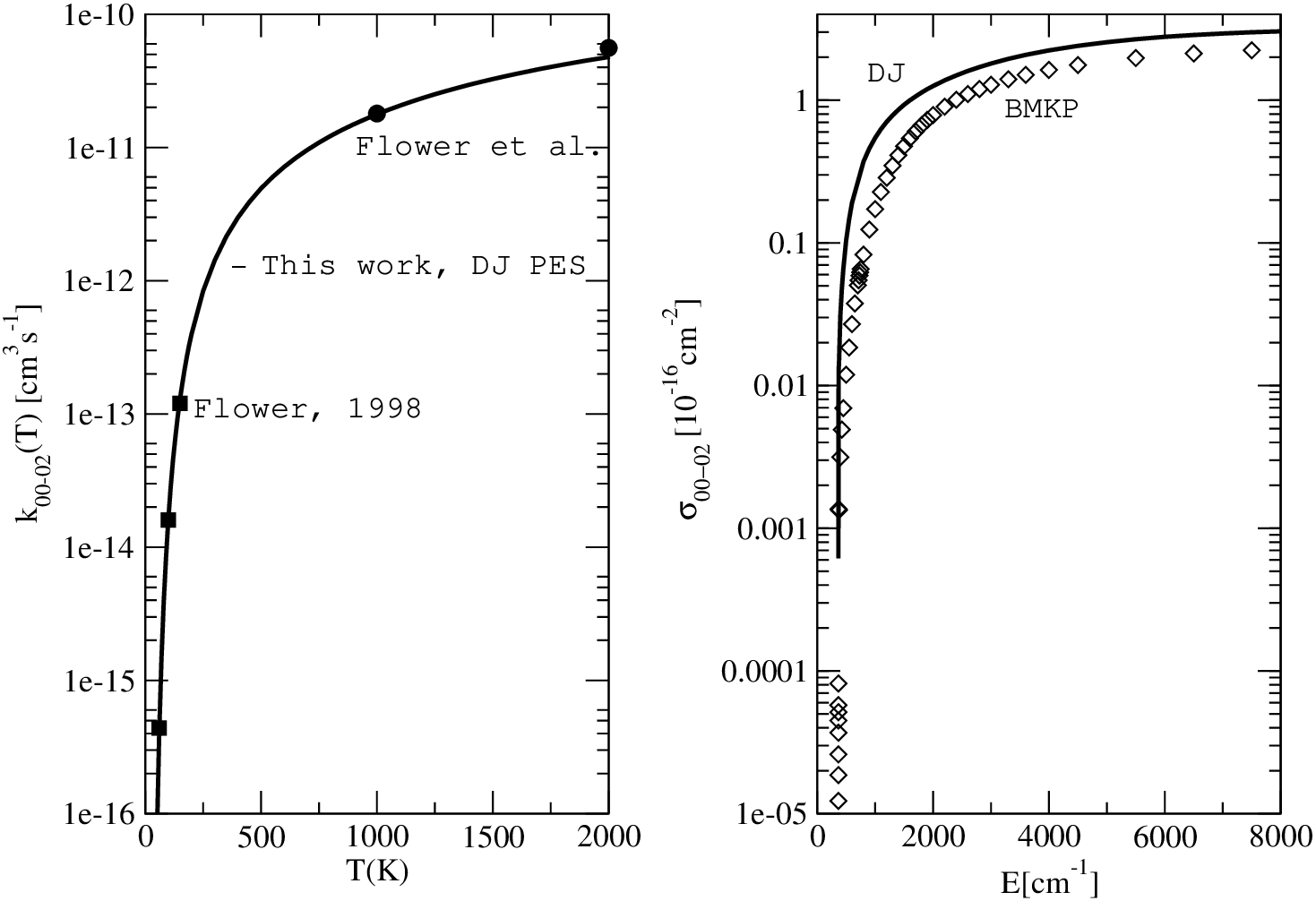}
\end{center}
\caption{Temperature dependence of the state-resolved thermal rate constant (left panel) and corresponding cross section
(right panel) for the transition $j_1=j_2=0 \rightarrow j'_1=2,j'_2=0$. Results from other works for the thermal rate
$k_{00-02}(T)$ are also included. The results for the DJ PES are given in solid lines.
The diamonds are the theoretical data of this work calculated with the BMKP PES.}
\label{fig:fig5}
\end{figure}

The BMKP PES \cite{booth02} is a global six-dimensional potential energy surface for two hydrogen molecules.
It was especially constructed to represent the whole interaction region of the chemical reaction dynamics of the 
four-atomic system and to provide an accurate as possible van der Waals well.
%
In the six-dimensional conformation space of the four atomic system the conical intersection forms a complicated 
three-dimensional hypersurface. The authors of the work \cite{booth02} mapped out a large portion of the locus 
of this conical intersection. 

The BMKP PES uses cartesian coordinates to compute distances between 
four atoms. We have devised some fortran code, which converts spherical coordinates used in
Sec. 2 to the corresponding cartesian coordinates and computes the distances between the four atoms. In all our
calculations with this potential the bond length was fixed at 1.449 a.u. or 0.7668 \r{A} as in DJ PES.


The main goal of this work is to carry out quantum-mechanical calculations for different transitions
in $p$-H$_2$+$p$-H$_2$ and $o$-H$_2$+$o$-H$_2$ collisions and to provide a comparative study of the two PESs 
presented above.
The energy dependence of the elastic integral cross sections $\sigma_{el} (E_{kin})$ are represented 
in Fig.\ 4 (upper plots) together with
the state-resolved integral cross sections $\sigma_{j_1j_2\rightarrow j'_1j'_2}(E_{kin})$
for the $j_1=j_2=0 \rightarrow j'_1=2,j'_2=2$ and $j_1=j_2=1 \rightarrow j'_1=1,j'_2=3$ rotational transitions
(lower plots)
for both the BMKP and DJ PESs respectively. As can be seen both PESs provide the
same type of the behaviour in the cross section. These results are in basic agreement with recent calculations,
but using a time-dependent quantum-mechanical approach \cite{guo02}.
Our calculation show, that DJ PES generates higher values for the cross sections.

A large number of test calculations have also been done to secure the convergence of the results with respect to all 
parameters that enter into the propagation of the Schr\"odinger equation. 
This includes the intermolecular distance $R$, the total angular momentum $J$ of the four atomic system, $N_{lvl}$ 
the number of rotational levels to be included in the close coupling expansion and others
(see the MOLSCAT manual \cite{hutson94}).

We reached convergence for the integral cross sections, $\sigma(E_{kin})$, in all considered collisions. In the case 
of DJ PES the propagation has been done from 2 \r{A} to 10 \r{A}, since this potential is defined only for those 
specific distances. 
For the BMKP PES we used $r_{min}=1$ \r{A} to $r_{max}=30$ \r{A}. We also applied a few different 
propagators included in the MOLSCAT program.


A convergence test with respect to the maximum value of
the total orbital momentum showed, that $J_{max}=100$ is good enough for the considered range of energies in this work.
%
We tested   
various rotational levels $j_1j_2$ included in the close coupling expansion
for the numerical propagation of the resulting coupled equations (\ref{eq:cpld}). 
In these test calculations we used two basis sets: $j_1j_2$=00, 20, 22, 40, 42 with total
basis set size $N_{lvl}=13$ and $j_1j_2$=00, 20, 22, 40, 42, 44, 60, 62 with $N_{lvl}=28$.
We found \cite{renat05a}, 
that the results are quite stable for the 00$\rightarrow$20 and 00$\rightarrow$22 transitions and somewhat
stable for the highly excited 00$\rightarrow$40 transition. Nontheless, for our production calculations we
used the first basis set.


It is important to point out
here, that for comparison purposes we don't include the compensating factor of 2 mentioned in \cite{flower87}.
However, in Fig.\ 4 (upper plots) and in our subsequent calculations of the thermal rate coefficients, $k_{jj'}(T)$,
the factor is included.

\begin{figure}
\begin{center}
\includegraphics[scale=1.0,width=23pc,height=21pc]{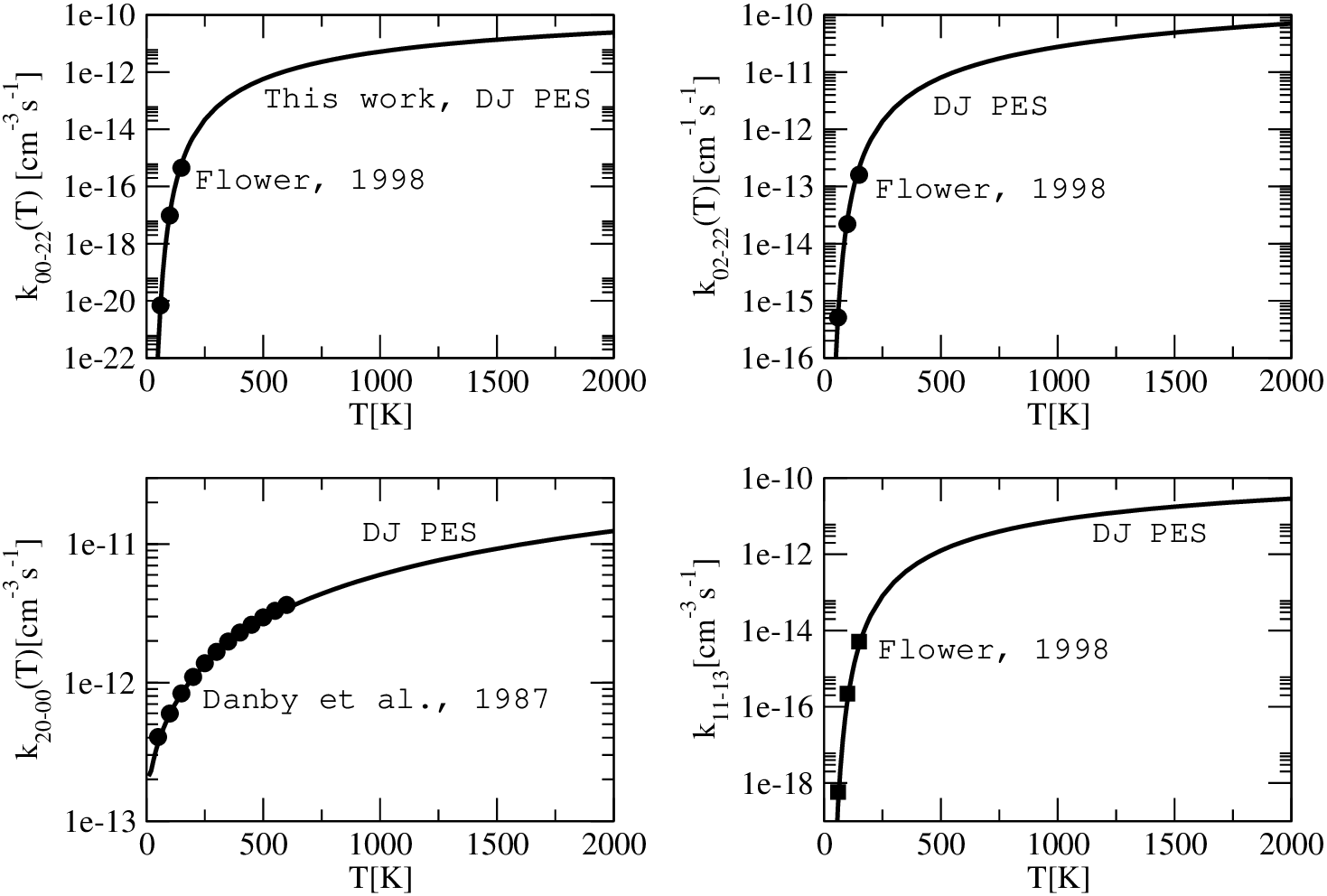}
\end{center}
\caption{Temperature dependence of the state-resolved thermal rate constants for the
$j_1=j_2=0 \rightarrow j'_1=2,j'_2=2$, $j_1=0,j_2=2 \rightarrow j'_1=2,j'_2=2$, $j_1=2,j_2=0 \rightarrow j'_1=0,j'_2=0$ and
$j_1=1,j_2=1 \rightarrow j'_1=1,j'_2=3$. Results obtained with the DJ PES.
Theoretical data from other works are included.}
\label{fig:fig6}
\end{figure}

\begin{figure}
\begin{center}
\includegraphics[scale=1.0,width=23pc,height=17pc]{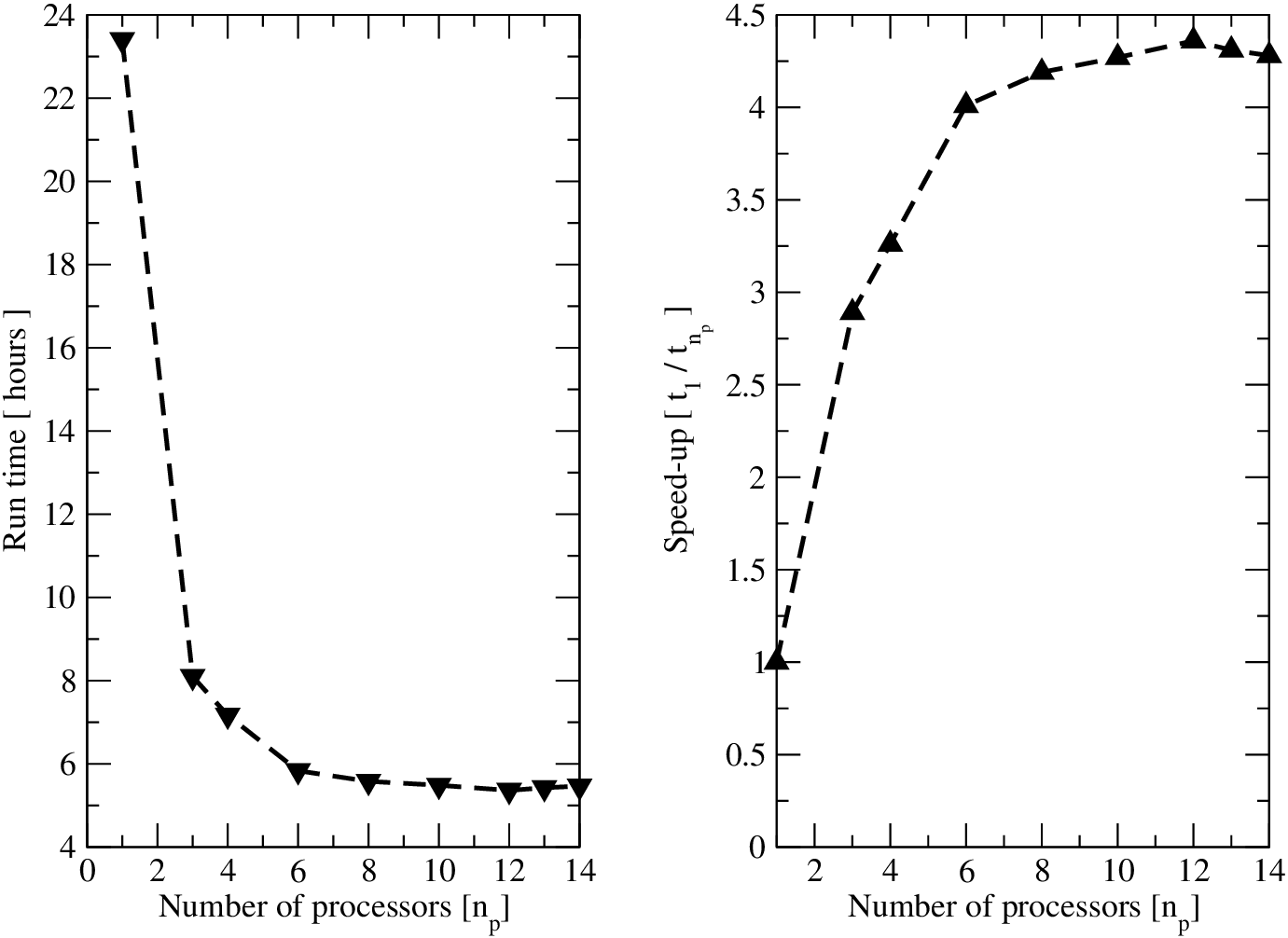}
\end{center}
\caption{Computation time and speed-up $t_1/t_{n_{p}}$ depending on number of parallel processors $n_p$ using the dynamic
approach, see text.}
\label{fig:fig7}
\end{figure}

The differences in the cross sections of the two potentials are reflected in the state-resolved
transition states $j_1=0,j_2=0 \rightarrow j'_1=0,j'_2=2$, as shown in Fig.\ 5 (right panel). 
It seems that the DJ PES can provide much better results, as seen in the same figure in the left panel, when we
present the results for the corresponding thermal rates $k_{00-02}(T)$ calculated with the DJ potential together with 
results of other theoretical calculations. The agreement is perfect. Thus, one can conclude, that
DJ PES is better suited for the H$_2-$H$_2$ system. In Fig.\ 6 we provide thermal rates for different transition
states calculated with only the DJ PES and in comparison with other theoretical data obtained within different
dynamical methods and PESs. Again the agreement is very good.

In Fig.\ 7 we present an example of
our timing results using the dynamic method for a specific H$_2(j_1)$+H$_2(j_2)$ calculations.
It can be seen, that including additional processors reduces the computation time.
Here we present two results. The left plot shows dependence of the computing time on amount of active 
parallel processors.
The right plot illustrates the degree of speed-up of the calculations. The speed-up for a fixed
test calculation is defined as $t_1/t_{n_{p}}$, where $t_1$ is the calculation with only one processor and 
$t_{n_p}$ with $n_p$ processors.

\section{Conclusion}


We carried out parallel computations for state-resolved rotational excitation and deexcitation cross sections and rates
in molecular $para$-/$para$- and ortho-/ortho-H$_2$ collisions of astrophysical interest. The LAM/MPI technique 
allowed us to speed up the computation process at least $\sim4.5$ times within our 14 processor Sun Unix cluster. We 
tested the two newest potential energy surfaces for the considered systems. Thus the application of the parallel algorithm 
reduced the computation time used to test the two potentials.
A test of convergence and the results for cross sections and rate coefficients
using two different potential energy surfaces for the H$_2-$H$_2$
system have been obtained for a wide range of kinetic energies.

We would like to point out here, that the hydrogen problem is very important for many reasons. The main motivation
has been described in the introduction of this paper. It is also necessary to stress, that the hydrogen-hydrogen
collision may be particularly interesting in nanotechnology applications, when the system is confined inside a single 
wall carbon nanotube (SWNT) \cite{gray03}.

Careful treatment of such collisions can bring useful information about the hydrogen adsorption mechanisms in SWNTs
and quantum sieving selectivities \cite{wang99}. However, in this problem particular attention should be paid not
only to the H$_2-$H$_2$ potential, but also to the many body interaction between H$_2$ molecules and the carbon 
nanotube [27-28]. The inclusion of additional complex potentials in the Schr\"odinger equation may
essentially increase the computation difficulties.

It is also very attractive to upgrade the four-dimensional model for the linear rigid rotors used in this work 
to complete six-dimensional consideration of the H$_2+$H$_2$ collisions. However,
because of two additional integrations over $r_1$ and $r_2$ distances such calculations should be very time consuming
\begin{eqnarray}
\left(\frac{d^2}{dR^2}-\frac{L(L+1)}{R^2}+k_{\alpha}^2\right)U_{\alpha}^{JM}(R)=2M_{12}
\sum_{\alpha'} \int \int <\phi^{JM}_{\alpha}(\hat r_1,\hat r_2,\vec R)\chi_{vj_1j_2}(r_1,r_2)\nonumber \\
|V(\vec r_1,\vec r_2,\vec R)| 
\chi_{v'j'_1j'_2}(r_1,r_2)\phi^{JM}_{\alpha'}(\hat r_1,\hat r_2,\vec R)>U_{\alpha'}^{JM}(R) 
d^3\vec r_1 d^3\vec r_2 d\hat R.
\label{eq:cpldnew}
\end{eqnarray}
Here $\chi_{v'j'_1j'_2}(r_1,r_2)$ is the product of the real vibrational wavefunctions of the two molecules
\begin{equation}
\chi_{vj_1j_2}(r_1,r_2)=w_{v_1j_1}(r_{1})w_{v_2j_2}(r_{2}),
\end{equation}
where $v$ designates the vibrational quantum numbers $v_1$ and $v_2$ \cite{depristo77}.
Nontheless, the application of a parallel computing techniques together with shared memory methodology could be a very 
effective computational approach, as it was partially demonstrated in this work.

Although our calculations revealed, that both the H$_2-$H$_2$ PESs used in this work
can provide the same type of behaviour in regard to cross sections and rates, there are still significant differences. 
%
%
Considering the results of these calculations we conclude that subsequent work is needed to further improve the 
H$_2-$H$_2$ PES, and that work will require parallel processing if it is to be done in a timely manner.


\vspace{5mm}


\section*{Acknowledgment}


This work was supported by the St. Cloud State University internal grant program, St. Cloud, MN (USA).



%

\end{document}